\documentclass{PoS}

\title{Tetraquarks in AdS/QCD}

\ShortTitle{Tetraquarks in AdS/QCD}

\author{\speaker{Hilmar Forkel}\\
        Institut f{\"u}r Physik, Humboldt-Universit{\"a}t zu Berlin, D-12489 Berlin, Germany\\
        E-mail: \email{forkel@physik.hu-berlin.de}}

\abstract{Multiquark correlations inside hadrons can have a significant
and in some cases even striking impact on the hadron spectrum. We 
show how such correlations in general, and mesons with a dominant 
tetraquark content in particular, emerge holographically in the AdS/QCD 
framework. On this basis, we arrive at a holographic realization of an 
exceptionally strong four-quark binding and a correspondingly large 
tetraquark component in the lightest scalar mesons. Higher-lying 
tetraquark excitations, on the other hand, become too broad to form 
supernumeral scalar states.}

\FullConference{Sixth International Conference on Quarks and Nuclear Physics\\
                 April 16-20, 2012\\
                 Ecole Polytechnique, Palaiseau,  Paris}

\begin{document}

\section{Introduction}

A dominant tetraquark component is often invoked to explain 
non-standard features of mesons ranging from the lightest scalar 
nonet \cite{jaf77} to some of the recently discovered heavy-quark 
resonances  \cite{god08}. Without guidance from the naive quark
model, the theoretical analysis of such exotic states has to deal
with the challenges of the strongly-coupled gauge dynamics 
more directly. Adding to the arsenal of suitable methods for this 
purpose (which now includes devoted lattice simulations 
\cite{mai07}), the AdS/QCD approach \cite{revs2} was recently 
extended to provide holographic insights into tetraquark
properties \cite{for10}.

The AdS/QCD program describes strong-interaction physics 
analytically by constructing approximate five-dimensional gravity 
duals for QCD on the basis of the gauge/string correspondence 
\cite{revs1}. To access multiquark correlations inside hadrons 
within this framework, one exploits the fact that such correlations 
manifest themselves in the quark composition of the QCD 
interpolator which most strongly couples to the hadron under
consideration. Hence these correlations 
leave a gauge-invariant imprint on the interpolator's anomalous 
dimension which the AdS/CFT dictionary \cite{revs1} translates into 
a multiquark-content-dependent mass correction for the hadron's
dual bulk mode. This holographic mechanism was previously employed 
to analyze the impact of  diquark correlations on masses and sizes 
of the light-quark baryons \cite{for09}. In the following, we will 
apply it to tetraquarks.

\section{Tetraquark holography}

The AdS/CFT dictionary  associates the bulk mode $\varphi(x,z) $ (where 
$x$ are the coordinates along the boundary and $z$ denotes the orthogonal
direction) dual to a given hadron $\left\vert h\right\rangle $ with the 
gauge-theory interpolator $J(x)$ which most strongly couples to 
$\left\vert h\right\rangle $. More specifically, for a scalar meson the scaling 
dimension $\Delta $ of $J$ determines the boundary condition \cite{revs1} 
\begin{equation}
\varphi \left( x,z\right) \stackrel{z\rightarrow 0}{\longrightarrow }\varphi
^{\left( 0\right) }\left( x\right) z^{4-\Delta }  \label{bc}
\end{equation}
for the dual bulk-mode solutions. This condition is imposed on the (at small $z$ 
leading) solutions of the bulk field equation by adjusting its mass term to
\begin{equation}
m_{5}^{2}R^{2}=\Delta \left( \Delta -4\right)  \label{m}
\end{equation}
(where $m_{5}$ is the mode's mass and $R$ is the AdS$_5$ curvature scalar).
Interpolators with a larger quark content and correspondingly larger 
$\Delta $ are therefore related to heavier bulk excitations and generally 
to mesons with larger mass. (Note that the relation (\ref{m}) is independent 
of the specific gravity dual under consideration).

In contrast to the unique quark-antiquark interpolator $J_{\bar{q}q}^{A}=
\bar{q}^{a}t^{A}q^{a}$ for ordinary scalar mesons with $\Delta _{\bar{q}q}=3$, 
there are several independent  (gauge-invariant and local) scalar four-quark 
operators with a common classical scaling dimension $\Delta _{\bar{q}^{2}q^{2}}=6$. 
(An explicit example is $J_{\bar{q}^{2}q^{2}}^{A}=\varepsilon ^{abc}\varepsilon ^{ade}\bar{q}^{b}C\Gamma ^{A}\bar{q}^{c}q^{d}C\Gamma ^{A}q^{e}$ which 
contains a ``good'' (i.e. maximally attractive) diquark and a good antidiquark.) 
Since all of these four-quark operators couple to the scalar tetraquark state 
$\left\vert t\right\rangle $, we define the (``optimal'') tetraquark 
interpolator $J_{\bar{q}^{2}q^{2}}$ as the particular linear combination which 
has maximal overlap $\left\langle 0\left\vert J_{\bar{q}^{2}q^{2}}
\right\vert t\right\rangle $ with the tetraquark ground state.

This definition encodes part of the tetraquark's structure into 
the field composition and anomalous dimension of $J_{\bar{q}^{2}q^{2}}$ 
since both are affected by the dominant quark couplings in 
$\left\vert t\right\rangle $. The anomalous dimension 
$\gamma\left( \mu \right)$, in particular, carries information on the 
multiquark correlations responsible e.g. for tetraquark binding \cite{for10}. 
Owing to the boundary condition (\ref{bc}) this information gets 
translated into a bulk mass correction $\Delta m_{5}\left( z\right)$ 
for the tetraquark's dual mode\footnote{For anomalous-dimension 
induced bulk-mass corrections in a different context see Ref. 
\cite{veg10}.}. (The $z$ dependence of $\Delta m_{5}$ is inherited 
from the dependence of $\gamma$ on the renormalization scale 
$\mu \sim 1/z$.) 

Despite their rather straightforward origin, the implementation 
of these bulk-mass corrections into bottom-up duals is not without 
challenges. To begin with, at present 
QCD information on anomalous dimensions of hadronic interpolators 
is still scarce (especially in the infrared) and their renormalization scheme 
dependence raises compatibility issues. A more fundamental concern  
\cite{for10} is the naive AdS/QCD extrapolation of the multiquark physics 
with its pronounced $N_{c}$ dependence to $N_{c}=3$. 
Major benefits of the outlined mechanism, on the other hand, include that it
will work in any of the current AdS/QCD duals, both non-dynamical (i.e. not
backreacted) and dynamical (see e.g.  Ref. \cite{dep09}), and that the results 
are essentially model independent. In the following we will implement this 
mechanism into the ``dilaton soft-wall'' dual of Ref.  \cite{kar06}
and apply it to the analysis of tetraquark correlations inside the lightest 
scalar mesons.

\section{Light scalar tetraquarks in the dilaton soft-wall dual}

The lightest scalar meson nonet (for reviews see e.g. \cite{clo02}) has long 
been suggested to contain a dominant tetraquark component \cite{jaf77}. 
This at present arguably favored interpretation requires an exceptionally 
strong four-quark binding which should be large enough to reduce the 
tetraquark mass considerably below that of the lightest scalar $\bar{q} q$ 
mesons \cite{veg08}. Hence light scalar tetraquarks furnish a particularly 
challenging testing ground for the holographic mechanism discussed above.

To implement the latter into an explicit dynamical framework, we adopt the 
soft-wall dual of Ref. \cite{kar06} which contains the AdS$_{5}$ 
metric and a quadratic dilaton field\footnote{in contrast to the 
"metric soft wall" of Ref. \cite{for07}} $\Phi \left( z\right) =\lambda ^{2}z^{2}$. 
After Fourier transforming the $x$ dependence of the tetraquark's dual modes 
$\varphi$ and defining a ``reduced'' scalar field $\phi\left( q,z\right)  = 
(R/z) ^{3/2}e^{-\lambda ^{2}z^{2}/2}\varphi\left( q,z\right)$, the radial 
bulk equation in this background turns into the Sturm-Liouville problem 
\begin{equation}
\left[ -\partial _{z}^{2}+V\left( z\right) \right] \phi \left( q,z\right)
=q^{2}\phi \left( q,z\right)  \label{sleq}
\end{equation}
with the potential 
\begin{equation}
V\left( z\right) =\left( \frac{15}{4}+m_{5}^{2}R^{2}\right) \frac{1}{z^{2}}
+\lambda ^{2}\left( \lambda ^{2}z^{2}+2\right) .  \label{vsw}
\end{equation}
For constant $m_{5}$ the normalizable solutions of this eigenvalue problem 
can be found analytically \cite{for10}. The resulting, discrete square-masses 
of the $n$-th radial excitations (related to $\Delta$ by Eq. (\ref{m})),
\begin{equation}
M_{n}^{2}=q_{n}^{2}=4 \lambda ^{2}\left( n+\frac{\Delta }{2}\right), \label{specs}
\end{equation}
lie on linear trajectories and indeed grow with $\Delta $. 

In addition, the spectra (\ref{specs}) reveal that the ground state of the 
four-quark trajectory with $\Delta _{\bar{q}^{2}q^{2}}=6$ has a twice larger 
square mass $M_{\Delta =6,0}^{2}=2M_{q\bar{q},0}^{2}$ than its 
$\Delta _{\bar{q}q}=3$ counterpart  (as long as anomalous dimensions 
are ignored). Simply replacing the conventional quark--antiquark interpolator 
by a $\bar{q}^{2}q^{2}$ interpolator would therefore result in four-quark 
states which are heavier, not lighter than the $\bar{q}q $ ground state. 
Of course, in view of our above discussion it is not surprising that the 
attraction required to bind tetraquarks is lacking. Indeed, if multiquark  
correlations generate the exceptional lightness of the tetraquark ground 
state, the latter should emerge from the so far neglected anomalous 
dimension of the tetraquark interpolator.

Inclusion of this anomalous dimension, which we denote by $\gamma(z)$, 
generalizes the bulk mass term (\ref{m}) to 
\begin{equation}
m_{5}^{2}\left( z\right) R^{2}=\left[ 6+\gamma \left( z\right) \right] \left[
2+\gamma \left( z\right) \right]  \label{m2}
\end{equation}
for the modes dual to the tetraquark interpolator $J_{\bar{q}^{2}q^{2}}$
with $\Delta _{\bar{q}^{2}q^{2}}=6+\gamma(z)$. This adds the universal 
correction 
\begin{equation}
\Delta V\left( z\right) =\gamma \left( z\right) \left[ \gamma \left(
z\right) +8\right] \frac{1}{z^{2}}  \label{DelV}
\end{equation}
to the potential (\ref{vsw}) with $m_{5}^{2}R^{2}=12$. Note that this correction
will be completely determined as soon as QCD information on the RG 
flow of $\gamma $ will become available. In the meantime, we resort 
to estimating its quantitative impact on the tetraquark spectrum and 
to deriving an exact lower bound on the ground-state mass. 
For the former purpose, we adopt the ansatz\footnote{Anomalous 
dimensions of a qualitatively similar $z$ dependence (in the region 
of interest) are encountered in dual backgrounds of holographic 
RG-flow type \cite{mue10}.} $\gamma \left( z\right) =-az^{\eta }+
bz^{\kappa }$ whose coefficients turn out to be tightly constrained 
by consistency and stability requirements \cite{for10}.

A fundamental property of Eq. (\ref{DelV}) is the lower bound 
$\Delta V\left(z\right) \geq -16/z^{2}$ which holds for any 
$\gamma $ and prevents the collapse of dual
modes into the AdS$_{5}$ boundary. This bound is saturated by 
$\gamma \equiv -4$ and therefore yields the lower bound
\begin{equation}
M_{\bar{q}^{2}q^{2},0}\geq M_{\Delta =2,0}=2\lambda  \label{mbd}
\end{equation}
which determines the lightest tetraquark mass which Eq. (\ref{m2}) can 
generate. Moreover, for constant values of $\gamma $ in the range $-4< 
\gamma <-3$ the tetraquark's ground-state mass remains below its 
$\bar{q} q $ counterpart. Reassuringly, the underlying binding mechanism 
will work in other AdS/QCD duals as well. Indeed, $\gamma $ enters the 
bulk dynamics exclusively through the mass term (\ref{m2}) 
which is model-independently prescribed by the AdS/CFT dictionary 
and yields the universal correction (\ref{DelV}) in any AdS/QCD dual.

\begin{figure}[tbp]
\includegraphics[width=.5\textwidth]{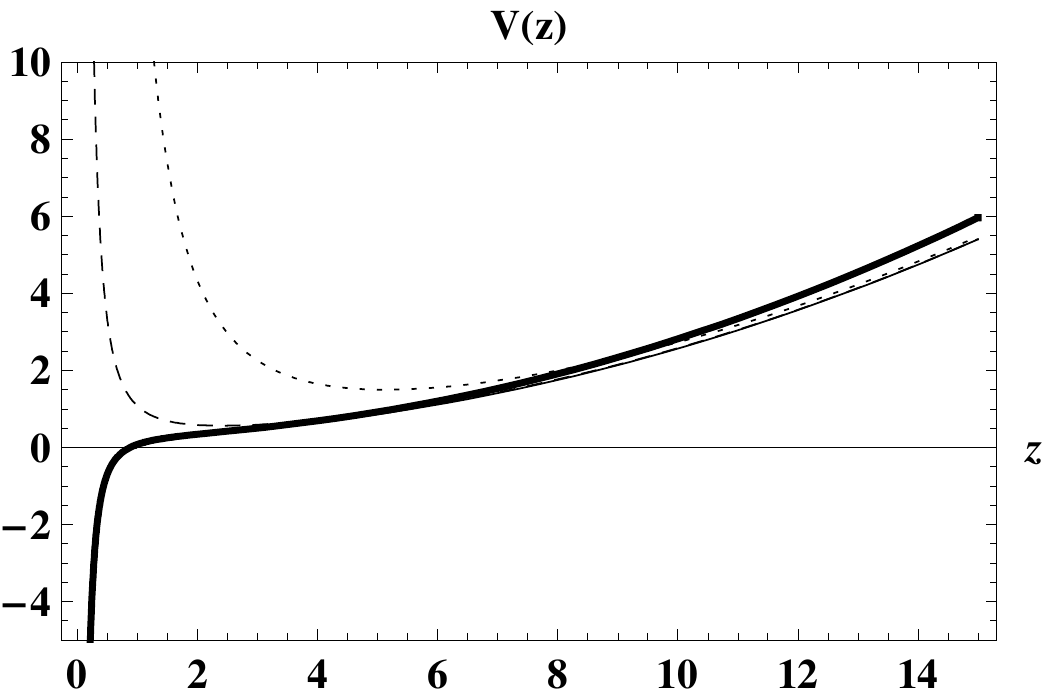} 
\includegraphics[width=.479
\textwidth]{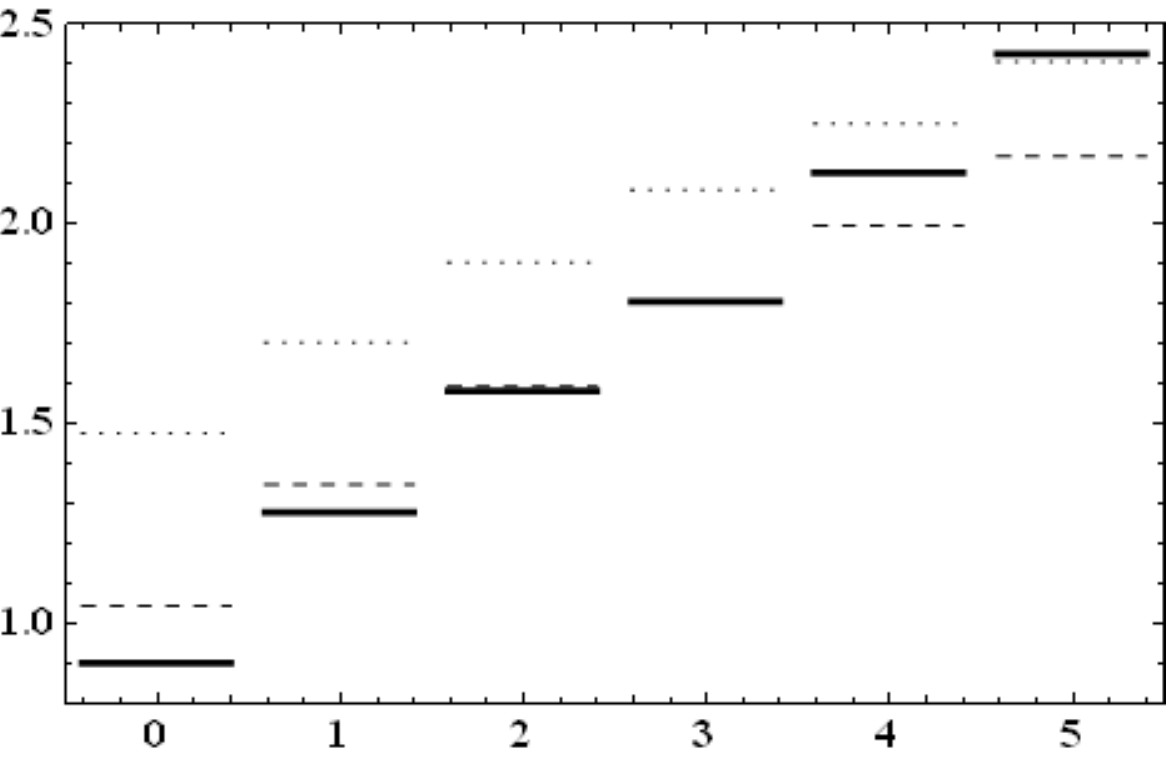}
\caption{Left panel: the tetraquark's dual-mode potential (\protect\ref{vsw})
(i) for the anomalous dimension $\gamma \left( z\right) =-az^{\eta }+
bz^{\kappa }$ with $a=4$, $b=0.05$, $\protect\eta =0.001$ and 
$\protect\kappa =2$ (thick line), (ii) for the four-quark interpolator with 
$\Delta =6,$ $\protect\gamma =0$ (dotted line), (iii) for the $\bar{q}q$ 
interpolator with $\Delta =3$ (dashed line) and (iv) for $\Delta =2$ 
(thin line) which saturates the Breitenlohner-Freedman bound.  Right 
panel: The ground state ($n=0$) and the first five excitations of the
tetraquark mass spectrum (i)  in the potential (i) of Fig. \protect\ref{fig1} 
(thick bars), (ii)  for the $\bar{q}q$ interpolator (dashed bars) and (iii) 
for the $\Delta =6$ interpolator (dotted bars) with $\protect\gamma =0$.}
\label{fig1}
\end{figure}

Higher-lying tetraquark excitations with masses $M_{\bar{q}^{2}q^{2},n}$ 
(for $n>0$)  \emph{beyond} those of the corresponding quark-antiquark
mesons require in addition a suitably running $\gamma \left( z\right) $. 
Our above ansatz, for example, can push the tetraquark
masses $M_{\bar{q}^{2}q^{2},n}$ beyond the $M_{\bar{q}
q,n}$ from around $n\gtrsim 2$ while simultaneously generating
an almost maximal ground-state binding energy. The latter drives 
$M_{\bar{q}^{2}q^{2},0}$ from $\sim 40\%$ above (for $\gamma 
\equiv 0$) down to $\sim 20\%$ below the $\bar{q}q$ ground-state 
mass\footnote{Hence the phenomenological ratio 
$M_{\bar{q}^{2}q^{2},0}/M_{\bar{q}q,0}\sim 0.8/1.5$ is not 
quantitatively reproduced, perhaps due to the neglect of the 
$\bar{q}q$ interpolator's anomalous dimension.} $M_{q\bar{q},0}
=\sqrt{6}\lambda $. 
This can be seen in Fig. \ref{fig1} where the corresponding spectra and 
potentials are plotted. It further appears that the higher-lying radial 
tetraquark excitations will be broad enough to prevent the appearance 
of supernumeral scalar states in the meson spectrum.

\section{Summary and conclusions}

\label{sum}

We have presented an essentially model-independent extension 
of the AdS/QCD framework which describes multiquark correlations 
in hadrons by means of anomalous-dimension-induced bulk-mass 
corrections. This mechanism yields a robust holographic description 
of a dominant tetraquark component in the lowest-lying scalar mesons. 
Moreover, it provides an explicit lower bound on holographically 
attainable tetraquark masses. In the dilaton soft-wall gravity dual this 
bound sits at twice the infrared scale and can be almost saturated. 
The resulting, exceptionally strong four-quark binding renders the tetraquark 
ground-state nonet about 20\% lighter than the lowest-lying scalar 
quark-antiquark state. Higher-lying tetraquark excitations, on the other 
hand, can be pushed beyond their quark-antiquark counterparts and are 
thus likely to dissolve into the multiparticle continuum.

\acknowledgments{It is a pleasure to thank the organizers of QNP 2012 for 
a very informative and enjoyable conference. Financial support from the 
Deutsche Forschungsgemeinschaft (DFG) is also acknowledged.}

\end{document}